\documentclass[prb,twocolumn,amsfonts]{revtex4}
\usepackage{graphicx}
\usepackage{amsmath}
\usepackage{amsfonts}
\usepackage{bm}

\def\ie{{\it {i.e.}}}
\def\eg{{\it {e.g.}}}

\begin{document}
\author{Samuel Marcovitch, Benni Reznik and Lev Vaidman }
\title{Quantum Mechanical Realization of a Popescu-Rohrlich Box}
\affiliation{ School of Physics and Astronomy,
Raymond and Beverly Sackler Faculty of Exact Sciences,
Tel-Aviv University, Tel-Aviv 69978, Israel.}

\date{\today}
\begin{abstract}

We consider quantum ensembles which are determined by
pre- and post-selection. Unlike the case of only pre-selected
ensembles, we show that in this case the probabilities for measurement outcomes
at intermediate times satisfy causality only rarely; 
such ensembles can in general be used to signal between causally disconnected regions.
We show that under restrictive conditions,  there are certain non-trivial bi-partite ensembles
which do satisfy causality.
These ensembles give rise to a violation of the CHSH inequality,
which exceeds the maximal quantum violation given by Tsirelson's bound,
$B_{\rm CHSH}\le 2\sqrt2$, and obtains 
the Popescu-Rohrlich bound for the maximal violation, $B_{\rm CHSH}\le 4$.
This may be regarded as an a posteriori realization of super-correlations,
which have recently been termed Popescu-Rohrlich boxes.

\end{abstract}
\maketitle
\section{Introduction}
One of the most remarkable features of quantum theory is the fact that it does not violate relativistic causality,
or specifically the no-signaling condition.
It seems that nothing in the formalism of quantum mechanics (QM) dictates causality.
Indeed, for quantum ensembles which are both
pre- and post-selected, the  probability law, given by Aharonov, Bergman and Lebowitz (ABL) \cite{abl},
does not in general satisfy causality.
Such ensembles can be used to signal between causally disconnected regions.

For a given pre- and post-selected ensemble, described by an initial state  $\vert \psi_i\rangle$
and a final state $\langle \psi_f\vert$, 
the probability of measuring state $\vert c_n\rangle$ at the intermediate time $t_i<t<t_f$ is given by \cite{abl,abl2}:
\begin{equation}
\label{bayes}
P(C=c_n\vert \psi_{i},\psi_{f})=
\frac{\vert \langle\psi_f(t)\vert P_{C=c_n}\vert \psi_{i}(t)\rangle\vert ^2}
{\Sigma_i\vert \langle\psi_f(t)\vert P_{C=c_i}\vert \psi_{i}(t)\rangle\vert ^2},
\end{equation}
where $P_{C=c_i}$ is a projection onto the space of eigenvalues equal to $c_i$.

Violation of causality can be shown \cite{aharonov vaidman}, for example, by taking the initial state to be a singlet
shared by Alice and Bob,
$\vert\psi_i\rangle=(\vert\! \uparrow_{z}\rangle\vert\!\downarrow_{z}\rangle-
\vert\!\downarrow_{z}\rangle\vert\!\uparrow_{z}\rangle)/\sqrt{2}$
and the final state to be
$\langle\psi_f\vert=\langle\uparrow_{x}\!\vert\langle\uparrow_{y}\!\vert$,
where the first particle belongs to Alice and the second to Bob.
(We will use this convention throughout the paper unless specified otherwise.)
If Bob  measures the spin component of his particle along the $x$ axis, 
then with certainty he will get $\vert\!\downarrow_{x}\rangle$.
However, if Alice also performs a measurement of her particle's spin along the $y$ axis,
Bob's probability for obtaining $\vert\downarrow_{x}\rangle$ reduces to 0.5.

The present paper focuses on the relations between causality and nonlocality in the context of pre- and post-selected ensembles.
In section \ref{sec2} we determine the generic classes of bi-partite pre- and post-selected ensembles that
satisfy causality.
We define causality in the context of the no-signaling condition. 
This condition imposes specific limitations on the allowed operations of the experimenters, 
which we shall explicitly define.
We prove that the ensembles of initial and final bi-partite spin half states that satisfy the no-signaling condition
belong to three generic classes.

In section \ref{sec3} we explore the amount of nonlocality in pre- and post-selected ensembles.
Popescu and Rohrlich (PR) \cite{pr} 
have already raised a similar question: can quantum nonlocality be derived from the no-signaling condition?
They discovered that it is possible to construct various causality satisfying models,
which exceed the quantum mechanical bound 
for the CHSH inequality \cite{chsh}, 
$B_{\rm CHSH}\le 2\sqrt2$, derived by Tsirelson \cite{cirelson}.
The maximal value of the CHSH inequality which satisfies causality is 4.
Such models that posses super-correlations, yet do not violate causality, have been termed PR-boxes 
and were elaborated in \cite{linden,b2,piani,massar}.
(Note that a similar analysis, with respect to \cite{pr}, 
has been conducted by Khalfin and Tsirelson in \cite{khalfin}).
Previous research \cite{van dam,short,cerf,buhrman} suggested theoretical applications, using these boxes, 
which cannot be implemented in QM. 
These include reduced communication complexity, 
bit-commitment and simulating projective measurements that can be performed on the singlet state without communication. 
However, it was found that 
the analogue of entanglement swapping \cite{swapping} cannot be implemented with these boxes \cite{short2}.

We show that for the non-trivial classes of causality satisfying initial and final states,
the CHSH inequality violation exceeds Tsirelson's bound and obtains the maximal value (4).
Thus these classes may be regarded as a posteriori realizations of PR boxes.
There is no classical analogue to the suggested scheme using pre- and post-selection.

In section \ref{sec4} we briefly discuss tri-partite systems. 
Then in section \ref{sec5} we discuss the implementation of the analogue of entanglement swapping \cite{swapping}
with pre- and post-selected ensembles.
Finally, in section \ref{sec6} we discuss the implementation of the suggested super-correlations in an experiment.  

\section{No-Signaling Condition}\label{sec2}
We begin by defining the no-signaling condition in the context of non-local boxes.
These boxes shall be taken here as an ensemble of pre- and post-selected states, defined by an initial state and a final state
on a pair of spin half particles.
No dynamics is introduced.
Two causally disconnected experimenters, Alice and Bob, can each ``ask'' the boxes a single question, \ie ,$\ $ 
perform a von-Neumann measurement on their single particle in an arbitrary direction, obtaining the probabilities
for the up and down outcomes.
The no-signaling condition requires that the choice of an experimenter whether to measure or not, and the 
direction of measurement if implemented, should not affect the other experimenter's probabilities.
This condition is a rather softened condition for causality
since the experimenters are neither allowed to perform POVM, nor to act on the states,
that is, to apply local unitary transformations.
If these restrictions are not imposed, only trivial ensembles (product initial and final states)
will satisfy causality.
We will elaborate on these restrictions shortly.
From now on causality implies the no-signaling condition defined above.
Note that generally, ``experimenters'' with PR boxes ask the boxes only single bit questions, 
which are characterized by $a\oplus b=x \cdot y$, where $x$ and $y$ are the input to the boxes (the questions) 
and $a$ and $b$ are the outputs.
This is less than is allowed in the present framework.

We now show that in bi-partite systems the no-signaling condition is satisfied for the following initial-final states:
\begin{enumerate}
\item Both product (trivial class):
$\vert\psi_i\rangle\!=\!\vert\!\!\uparrow\rangle\vert\!\!\uparrow'\rangle,\quad
\langle\psi_f\vert=\langle\uparrow''\vert\langle \uparrow'''\vert,$ 
where the tag(s) denote different bases.
\item Both maximally entangled:
\begin{eqnarray}
\label{max}
&&\vert \psi_i\rangle=\frac{1}{\sqrt{2}}\big(\vert \uparrow\rangle\vert \uparrow'\rangle+e^{i \theta_i}\vert 
\downarrow\rangle\vert \downarrow'\rangle\big),
\nonumber
\\&&\langle \psi_f\vert=\frac{1}{\sqrt{2}}\big(\langle \uparrow''\vert\langle \uparrow'''\vert+
e^{-i \theta_f}\langle \downarrow''\vert\langle \downarrow'''\vert\big).
\end{eqnarray}
\item Equal states, but with their amplitudes swapped:
\begin{eqnarray}
\label{swap}
&&\vert \psi_i\rangle=\sqrt{\alpha}\lvert \uparrow\rangle\vert \uparrow'\rangle+
e^{i \theta}\sqrt{1-\alpha}\vert \downarrow\rangle\vert \downarrow'\rangle,
\nonumber
\\
\nonumber
\\&&\langle \psi_f\vert=\sqrt{1-\alpha}\langle\uparrow\!\vert\langle \uparrow'\vert+
e^{-i \theta}\sqrt{\alpha}\langle \downarrow\vert\langle \downarrow'\vert.
\nonumber
\\ \scriptstyle
\end{eqnarray}
\end{enumerate}
This entails that generally, even if the amount of entanglement in the initial and final states is the same, causality may be violated.

For simplicity we choose the Schmidt decomposition basis $\{\vert \uparrow_{}\rangle\vert\uparrow_{}\rangle$,
$\vert \downarrow_{}\rangle\vert\downarrow_{}\rangle\}$ for $\vert\psi_i\rangle$ and
$\{\langle \tilde\uparrow_{}\vert\langle\tilde\uparrow_{}\vert,
\ \langle \tilde\downarrow_{}\vert\langle\tilde\downarrow_{}\vert\}$ for $\langle\psi_f\vert$:
\begin{equation}
\begin{split}
&\vert\psi_i\rangle=\sqrt{\alpha}\vert \uparrow_{}\rangle\vert \uparrow_{}\rangle+
e^{i\theta_i}\sqrt{1-\alpha}\vert \downarrow_{}\rangle\vert \downarrow_{}\rangle,
\\&\langle\psi_f\vert=\sqrt{\beta}\langle \tilde\uparrow_{}\vert\langle \tilde\uparrow_{}\vert+
e^{-i\theta_f}\sqrt{1-\beta}\langle\tilde\downarrow_{}\vert\langle \tilde\downarrow_{}\vert,
\end{split}
\nonumber
\end{equation}
where $0\leq\alpha,\beta\leq1$.
Alice can freely choose the direction of her measurement. It can therefore be written as the projection operator
$P_{A\uparrow}=V_A\vert\!\!\uparrow\rangle\langle\uparrow\!\!\vert V_A^\dagger$
used in (\ref{bayes}). Correspondingly, Bob's projection operator is
$P_{B\uparrow}=V_B\vert\!\!\uparrow\rangle\langle\uparrow\!\!\vert V_B^\dagger$.
Unitary transformations $V_A$ and $V_B$ rotate the spins and are represented by
\begin{equation}
\label{u}
V_{A,B}=\begin{pmatrix}\cos{(\omega_{A,B}/2)}&-e^{-i\phi_{A,B}}\sin{(\omega_{A,B}/2)}\\e^{i\phi_{A,B}}
\sin{(\omega_{A,B}/2)}&\cos{(\omega_{A,B}/2)}
\end{pmatrix}.
\end{equation}
The no-signaling condition for Alice can be written as follows:
\begin{eqnarray}
&&P_A(i)=P_{AB}(i,j)+P_{AB}(i,\tilde j),
\label{cona1}
\\&&P_{AB}(i,j)+P_{AB}(i,\tilde j)=P_{AB}(i,j')+P_{AB}(i,\tilde j'),\ \ 
\label{cona}
\end{eqnarray}
where $P_{AB}(i,j)$ is the joint probability of obtaining $A=i$ and $B=j$
when both $A$ and $B$ are measured, and 
$\langle i\vert \tilde i \rangle=\langle j\vert \tilde j 
\rangle=\langle i'\vert \tilde i' \rangle=\langle j'\vert \tilde j' \rangle= 0$.
Eq. (\ref{cona1}) implies that Alice's probability does not depend on Bob's choice 
of whether to measure or not,
while (\ref{cona}) implies that Alice's probability is independent of Bob's choice of direction.
An analogue condition holds for Bob.
We immediately see that (\ref{cona}) is contained in (\ref{cona1}),
since (\ref{cona1}) does not only require that the terms be equal, but specifically determines their value.
From the ABL rule (\ref{bayes}):
$$P_{AB}(i,j)+P_{AB}(i,\tilde j)=\frac{\hat p_{ij}+p_{i\tilde j}\hat p_{i\tilde j}}{p_{ij}\hat p_{ij}+
p_{i\tilde j}\hat p_{i\tilde j}+p_{\tilde ij}\hat p_{\tilde ij}+p_{\tilde i\tilde j}\hat p_{\tilde i\tilde j}},$$
where $p_{ij}$ is the standard quantum mechanical probability for Alice obtaining $i$ and Bob $j$ when measuring 
$\vert\psi_i\rangle$, $p_{ij}=\langle\psi_i\vert ij\rangle\langle ij\vert \psi_i\rangle$.
$\hat p_{ij}$  is the corresponding probability for $\langle\psi_f|$, 
$\hat p_{ij}=\langle\psi_f\vert ij\rangle\langle ij\vert \psi_f\rangle$.
$P_A(i)$ can be expanded as
\begin{equation}
\begin{split}
&P_A(i)=\frac{\vert \langle\psi_f\vert ij\rangle\langle ij\vert \psi_i\rangle+
\langle\psi_f\vert i\tilde j\rangle\langle i\tilde j\vert \psi_i\rangle\vert ^2}
{\Sigma_{i\in \{i,\tilde i\}}\vert \langle\psi_f\vert ij\rangle\langle ij\vert \psi_i\rangle+
\langle\psi_f\vert i\tilde j\rangle\langle i\tilde j\vert \psi_i\rangle\vert ^2}
\\&=\frac{p_{ij}\hat p_{ij}+p_{i\tilde j}\hat p_{i\tilde j}+a(ij,i\tilde j)}
{p_{ij}\hat p_{ij}+p_{i\tilde j}\hat p_{i\tilde j}+p_{\tilde ij}\hat p_{\tilde ij}+p_{\tilde i\tilde j}\hat p_{\tilde i\tilde j}+
a(ij,i\tilde j)+a(\tilde ij,\tilde i\tilde j)},
\end{split}
\nonumber
\end{equation}
where
\begin{equation}
\begin{split}
a(ij,i\tilde j)&=\langle\psi_f\vert ij\rangle\langle ij\vert \psi_i\rangle
\langle\psi_i\vert i\tilde j\rangle\langle i\tilde j\vert \psi_f\rangle+c.c. 
\\&=2\sqrt{p_{ij}  \hat p_{ij} p_{i\tilde j}\hat p_{i\tilde j}} 
\cos(\alpha_{ij}-\hat\alpha_{ij}-\alpha_{i\tilde j}+\hat\alpha_{i\tilde j}).
\nonumber
\end{split}
\end{equation}
$\alpha_{ij}$ is the argument of the complex amplitude $\langle ij\vert \psi_i\rangle$, 
while $\hat\alpha_{ij}$ is the argument of $\langle ij\vert \psi_f\rangle$.
To simplify, we denote $ij$ as 1, $i\tilde j$ as 2, $\tilde i j$ as 3, and $\tilde i\tilde j$ as 4.
Note that every probability for a measurement outcome performed on $\vert\psi_i\rangle$ 
is multiplied by the corresponding probability of
$\langle\psi_f|$. 
We therefore denote
$\ p_{ij}\hat p_{ij}$ by $p_1$, $\alpha_{ij}-\hat\alpha_{ij}$ by $\alpha_{1}$, 
$\ p_{i\tilde j}\hat p_{i\tilde j}$ by $p_2$, etc.
The no-signaling condition can now be expressed as:
\begin{equation}
\label{cond}
\begin{split}
&(p_1\!+\!p_2)\sqrt{p_3 p_4}\cos(\alpha_3\!-\!\alpha_4)\!=\!(p_3\!+\!p_4)\sqrt{p_1 p_2}\cos(\alpha_1\!-\!\alpha_2),
\\&(p_1\!+\!p_3)\sqrt{p_2 p_4}\cos(\alpha_2\!-\!\alpha_4)\!=\!(p_2\!+\!p_4)\sqrt{p_1 p_3}\cos(\alpha_1\!-\!\alpha_3),
\end{split}
\end{equation}
for all bases $i,j$.
Since the equations have a symmetric form,
the general solutions for these conditions are:
\begin{eqnarray}
(p_1+p_2)\sqrt{p_3 p_4}&=&(p_3+p_4)\sqrt{p_1 p_2},
\nonumber
\\(p_1+p_3)\sqrt{p_2 p_4}&=&(p_2+p_4)\sqrt{p_1 p_3},
\nonumber
\\\cos(\alpha_3-\alpha_4)&=&\cos(\alpha_1-\alpha_2),
\nonumber
\\\cos(\alpha_2-\alpha_4)&=&\cos(\alpha_1-\alpha_3),
\nonumber
\end{eqnarray}
yielding:
\begin{eqnarray}
\label{cond2}
&&(p_3-p_2)(p_1+p_4)(p_1 p_4-p_2 p_3)=0,
\nonumber
\\&&\alpha_1+\alpha_4=\alpha_2+\alpha_3\  \textrm{or} \  \alpha_2=\alpha_3,\ \alpha_1=\alpha_4.
\end{eqnarray}

It is now possible to identify the causality satisfying states.
First, it can easily be shown that it is only for product states that $p_{ij}p_{\tilde i\tilde j}=p_{i\tilde j}p_{\tilde i j}$ and
$\alpha_{ij}+\alpha_{\tilde i\tilde j}=\alpha_{i\tilde j}+\alpha_{\tilde i j}$ for all bases $i,j$.
It follows that only for initial and final product states $p_1 p_4=p_2 p_3$ and $\alpha_1+\alpha_4=\alpha_2+\alpha_3$.

In addition, $p_2-p_3=0$ only if $p_{i\tilde j}=p_{\tilde i j},\ \hat p_{i\tilde j}=\hat p_{\tilde i j}$ or
$p_{i\tilde j}=\hat p_{\tilde i j},\ \hat p_{i\tilde j}=p_{\tilde i j}$.
The first condition corresponds to the maximally entangled class 
(\ref{max}) and the second corresponds to the swapped class (\ref{swap}).
For both classes $p_1=p_4$.
For maximally entangled states $\alpha_{ij}=\alpha_{\tilde i\tilde j}$ and $\alpha_{i\tilde j}=\alpha_{\tilde i j}$,
so that $\alpha_1=\alpha_4$ and $\alpha_2=\alpha_3$.
For the swapped class:
$\alpha_{ij}=\hat\alpha_{\tilde i\tilde j},\ \alpha_{\tilde i\tilde j}=\hat\alpha_{ij},
\ \alpha_{i\tilde j}=\hat\alpha_{\tilde i j}$ and $\alpha_{\tilde i j}=\hat\alpha_{i\tilde j}$,
so that, $\alpha_1+\alpha_4=\alpha_2+\alpha_3$. 
Each of the two classes, therefore, satisfy the second line in (\ref{cond2}).
In addition, both classes satisfy
\begin{equation}
\label{half}
P_A(i)=\frac{p_1+p_2+d_{12}}
{p_1+p_2+p_3+p_4+d_{12}+d_{34}}=\frac{1}{2},
\end{equation}
where $d_{nm}=2\sqrt{p_n p_m}\cos(\alpha_n-\alpha_m)$.
Thus, whenever only Alice or Bob implements a measurement, each outcome probability equals half.

The last possibility, $p_1=p_4=0$, does not yield any states since there is no state for which two of the probabilities are zero for all
measurement directions.
Note that in general even equal initial and final states do not satisfy causality.
Finally, we remark that if the post-selected state is known to be maximally entangled, yet is otherwise unknown, 
then causality is trivially satisfied. 
We therefore recover standard QM with a  Bell measurement applied to the state.

We can now discuss why Alice and Bob may only measure their states in a von-Neumann manner 
and may not apply any unitary transformations.
For otherwise, all non trivial causality satisfying states
would enable signaling. 
First, let us assume the ensemble contains maximally entangled initial and final states, for example, both singlet states.
Then Alice can measure the spin of her particle along the $z$ direction and flip it only if it is found to be down, 
restricting it to the up state. 
However, since the state is post-selected to be a singlet, 
Bob's state is surely down, if measured along the $z$ direction.
But if Alice had not applied the conditional flip, Bob would have measured both states with equal probability.
Second, for the swap class the situation is even worse. 
Here Alice may only implement a unitary transformation rotating the spin of her particle,
without measuring.
This transformation changes the pre-selected state to a different state, specifically not the swap state, 
which changes Bob's measurement outcomes probabilities.
Hence, local transformations rearrange the pre- and post-selected ensembles and therefore generally enable signaling.
Consequently, extending von-Neumann measurements to POVMs allows the implementation of conditional unitary transformations
(for example by using ancillas), and for this reason are excluded as well.


\section{Exceeding Tsirelson's Bound}\label{sec3}

We proceed by deriving the bound on the CHSH inequality \cite{chsh} for the non-trivial causality satisfying ensembles 
(\ref{max}, \ref{swap}).
The CHSH inequality (which holds in any classical local realistic theory) states that a certain combination of correlations is bounded,
$B_{\rm CHSH}=\vert C(A,B)-C(A,B')+C(A',B)+C(A',B')\vert \leq2$,
where the observables $A$, $A'$, $B$ and $B'$ take the values of $\pm1$, and the correlation $C(A,B)$ is defined as
$P_{AB}(1,1)+P_{AB}(-1,-1)-P_{AB}(1,-1)-P_{AB}(-1,1)$.
In a classical theory any observable has a predefined value and the inequality is satisfied trivially.
However, quantum correlations for bi-partite systems violate the CHSH inequality
and may reach Tsirelson's bound $B_{\rm CHSH}\le 2\sqrt2$ \cite{cirelson}. 

Returning to the pre- and post-selected ensembles we first consider bi-partite ensembles with 
a maximally entangled initial and final state (\ref{max}).
In this class, the CHSH inequality may achieve the maximal value $B_{\rm CHSH}=4$,
\eg $\ $ for the initial and final states:
\begin{eqnarray}
\label{maximal}
&&\vert\psi_i\rangle=\frac{1}{\sqrt{2}}\big(\vert\uparrow_{z}\rangle\vert \uparrow_{z}\rangle+
\vert \downarrow_{z}\rangle\vert \downarrow_{z}\rangle\big),
\nonumber
\\&&\langle\psi_{f}\vert=\frac{1}{\sqrt{2}}\big(\langle \uparrow_{z}\vert\langle \uparrow_{x}\vert-
\langle \downarrow_{z}\vert\langle \downarrow_{x}\vert\big).
\end{eqnarray}
If Alice and Bob perform measurements along the $z$ and $x$ axes, they yield the correlations:
$C(Z_A,Z_B)=C(X_A,X_B)=C(Z_A,X_B)=1$, and $C(X_A,Z_B)=-1$.
Therefore $B_{\rm CHSH}=\vert C(Z_A,Z_B)-C(X_A,Z_B)+C(Z_A,X_B)+C(X_A,X_A)\vert= 4$.
Thus the condition $a\oplus b=x \cdot y$ is satisfied, where $x$ and $y$ are the inputs to the boxes,
$x\in\{X_A=1,\ Z_A=0\}$,\  $y\in\{X_B=0,\ Z_B=1\}$, and 
$a$ and $b$ are the output of the boxes, 
and we take spin up to be $1$ and spin down to be $0$. 

Now let us examine the maximal bound on the CHSH inequality for the third causality satisfying class --
the swapped class (\ref{swap}).
Here the CHSH inequality may only saturate the maximal bound of $4$.
Interestingly, as the entanglement of the swapped states increases, the maximal achievable bound decreases.
In the extremal state in which both states are maximally entangled and equal: 
$$|\psi_i\rangle=|\psi_f\rangle=\frac{1}{\sqrt{2}}(|\uparrow_z\rangle|\uparrow_z\rangle+
e^{i\theta}|\downarrow_z\rangle|\downarrow_z\rangle),$$
one finds that $B_{\rm CHSH}\leq8\sqrt{2}/3$, which is the minimal value for the swapped states. 
As the amplitudes differ and the entanglement is reduced, the maximal bound increases.
However, when the amplitudes equal $1$, corresponding to product initial and final states,
the maximal bound jumps to 2.
The correlation of two observables $A$,$B$ measured by Alice and Bob is given by
\begin{widetext}
\begin{equation}
\label{long}
C(A,B)\!=\!\frac{16x\cos(\omega_A)\cos(\omega_B)+8\sqrt{x}\cos(\phi-\theta)
\sin(\omega_A)\sin(\omega_B)}
{x\big(3\!+\!\cos(2\omega_A)\big)\big(3\!+\!\cos(2\omega_B)\!\big)\!+\!2\big(1\!+\!2x\cos(2\phi\!-\!2\theta)\!\big)
\sin^2(\omega_A)\sin^2(\omega_B)\!+2\!\sqrt{x}\cos(\phi\!-\!\theta)\sin(2\omega_A)\sin(2\omega_B)}
\end{equation}
\end{widetext}
where $x=\alpha-\alpha^2$, $\phi=\phi_A+\phi_B$ and 
$\omega_A$, $\phi_A$, $\omega_B$ and $\phi_B$ are the measurement directions chosen by Alice and Bob respectively, 
as described in (\ref{u}).
$C(A,B)$ equals $1$ for $\omega_A=\omega_B=0$.
We define the measurement directions:
$\omega_A,\  \omega_A',\  \phi_A$ and $\phi_A'$ corresponding to $A$ and $A'$ 
and $\omega_B,\  \omega_B',\  \phi_B$ and $\phi_B'$ similarly for $B$ and $B'$.
In order to find maximal bound on the CHSH inequality, one can choose $\phi_A=\phi_A'$, $\phi_B=\phi_B'$ and $\phi_A+\phi_B=\theta$.
It can then be shown that for
$\omega_A=3\pi/2,\ \omega_A'=\pi$, $\omega_B=\pi+\pi d(\alpha)/4$, and $\omega_B'=\pi-\pi d(\alpha)/4$,
one obtains the maximal value of the CHSH inequality for a given $\alpha$, where $d(\alpha)$ is found numerically.
$d(\alpha\!=\!1/2)\!=\!1$ reduces monotonically as $\alpha$ increases (or decreases), 
while $d(\alpha\rightarrow 0\ \textrm{or} \ 1)\rightarrow 0$.
The last term, $d(\alpha\rightarrow 0\ \textrm{or} \ 1)\rightarrow 0$, 
corresponds to infinitesimal entanglement in which  $B_{\rm CHSH}\rightarrow4$.
Already for $\alpha=0.2$, where $d(\alpha\!=\!0.2)\!=\!0.505$, the maximal bound on the CHSH inequality is 
$B_{\rm CHSH}\approx 3.993$.


\section{Tri-Partite Systems}\label{sec4}
We now briefly discuss the no-signaling condition for initial and final tri-partite states.
Here the no-signaling condition becomes even worse --- even with maximally entangled initial and final states such as the GHZ states
(shared by Alice, Bob and Clare), signaling is possible. Take for example:
\begin{equation}
\begin{split}
&\vert \psi_i\rangle=\frac{1}{\sqrt{2}}\big(\vert \uparrow_{y}\rangle\vert \uparrow_{y}\rangle\vert \uparrow_{y}\rangle+
\vert \downarrow_{y}\rangle\vert \downarrow_{y}\rangle\vert \downarrow_{y}\rangle\big),
\\&\langle \psi_f\vert=\frac{1}{\sqrt{2}}\big(\langle \uparrow_{x}\vert\langle \uparrow_{x}\vert\langle \uparrow_{x}\vert-
\langle \downarrow_{x}\vert\langle \downarrow_{x}\vert\langle \downarrow_{x}\vert\big).
\\
\end{split}
\end{equation}
If Alice chooses to measure her spin along the $z$ direction, then $P_A(\downarrow)=0$,
while if Bob implements a measurement along the $x$ direction, then $P_A(\uparrow)=0.5$.
We estimate that only the trivial initial and final product states generally satisfy causality.

\section{Entanglement Swapping}\label{sec5}
Recently, Short, Popescu and Gisin \cite{short2} showed that it is 
impossible to implement the analogue of entanglement swapping \cite{swapping} with PR boxes.
Assume that Alice and Bob share a PR box 
and Bob and Clare share a PR box too. 
It was found that Bob cannot swap the correlations, that is, he cannot create any nonlocal correlations between Alice and Clare.

In our case, if we limit Bob to perform only single particle measurements, in analogue with \cite{short2}, 
then the same conclusion is reaffirmed:
no non-local correlations can be created between Alice and Clare.
However, if we allow Bob to perform any operations on his particles, 
then he can create a PR box between Alice and Clare, 
at least if the PR boxes are 
constructed from maximally entangled initial and final states
as in (\ref{maximal}).

The method to achieve this is similar to entanglement swapping in standard QM \cite{swapping},
as it is also based on Bell measurements.
Alice and Bob and Clare and Bob share an ensemble of pre- and post-selected states
in the form of (\ref{maximal}).
In order to perform swapping, Bob performs Bell measurements on his particles
and the Hadamard operation 
\begin{equation}
H=\frac{1}{\sqrt{2}}\left( \begin{array}{rr}1 & 1\\-1 & 1\end{array} \right)
\nonumber
\end{equation}
on the particle that he shares with Clare.
If Bob obtains 
$|\phi^+\rangle=\big(|\uparrow_z\rangle|\uparrow_z\rangle+|\downarrow_z\rangle|\downarrow_z\rangle\big)/\sqrt{2}$, 
then Alice and Clare share an ensemble of pre-and post-selected states in the form of (\ref{maximal}), which is a PR box.
If Bob obtains other outcomes in his Bell measurements, then we still get pre- and post-selected ensembles that
possess maximal correlations, but with other variables.
Thus Alice and Clare obtain a PR box only after Bob transmits the classical information of the measurement outcome.

It should be mentioned though, that if we let Bob perform
any bi-partite measurements on his particles, in general, he will also create signaling between Alice and Clare.
After a measurement in a basis of states that are entangled, but not maximally entangled,
the states shared by Alice and Clare do not belong to any of the classes that satisfy 
the no-signaling condition, which were found in section \ref{sec2}.
Therefore, the moment the information of Bob's observed outcome reaches Alice and Clare, 
they can superluminally signal each other.

\section{Physical Realization}\label{sec6}
We proceed now by suggesting a physical realization of 
bi-partite super-correlations using the maximally entangled initial and final states (\ref{maximal}). 
The scheme to demonstrate such
correlations in an experiment includes three steps.
First the desired EPR state is prepared.
The next step is a simulation of the
super-correlations by implementation of local measurements
for each site in the $x$ or $z$ directions.
Finally, a measurement is made to verify that the correct EPR
state has been obtained.
This happens in a quarter of the cases.
In which the intermediate correlations
yield $B_{CHSH}=4$.
The most practical implementations of such
experiments can be realized with photons. Preparation and verification of
EPR states with photons have been conducted in teleportation experiments 
\cite{photon}.
The intermediate measurements are implemented with polarization filters
in the desired orientations.
However, in such experimental setups,
there will be no clicks in the detectors in the intermediate measurements and
their success or failure is given a posteriori.
A more illustrative yet difficult to
implement experiment can be conducted with ion traps.
Preparation and verification of EPR states
with ion traps have recently been conducted \cite{ion}.
It should be mentioned that the obtained maximal correlations
require  communication in the post-selection procedures.

\section{Discussion}
Clearly, Quantum mechanics satisfy causality.
However, variants of the theory, such as nonlinear dynamical theories \cite{gisin}, generally violate causality. 
The existence of a final state and its relation to causality in the context of a universal wavefunction 
have been discussed in \cite{gruss,gell-mann,kent}.
In the present paper we showed that though an additional boundary condition 
generally leads to causality violation,
one can define natural constraints of single particle measurements, for which there are non trivial pre- and post-selected states 
that satisfy causality.
These pre- and post-selected states
give rise to a violation of the CHSH inequality, which exceeds the regular
quantum mechanical bound and reaches the maximal value of 4. 
Cabello \cite{cabello} has proposed to reach this bound using post-selection on GHZ states.
Our method provides an a posteriori PR box which can be implemented with today's technology.

\section{Acknowledgments}
We thank Y. Aharonov, J. Silman, J. Kupferman and M. Marcovitch for helpful discussions.
This work has been supported by the European Commission under the
Integrated Project Qubit Applications (QAP) funded by the IST
directorate as Contract Number 015848.


\end{document}